# Introducing a New Brexit-Related Uncertainty Index: Its Evolution and Economic Consequences

**Dr. Ismet Gocer***

Southampton Solent University
Southampton, UK.
ORCID: 0000-0001-6050-1745
ismet.gocer@solent.ac.uk
* Corresponding Author

**Dr. Julia Darby**

University of Strathclyde,
Glasgow, UK.
0000-0003-4425-7222
julia.darby@strath.ac.uk

**Dr. Serdar Ongan**

University of South Florida
Tampa, USA
0000-0001-9695-3188
serdar.ongan@gmail.com

**Abstract**

Important game-changer economic events and transformations cause uncertainties that may affect investment decisions, capital flows, international trade, and macroeconomic variables. One such major transformation is Brexit, which refers to the multiyear process through which the UK withdrew from the EU. This study develops and uses a new Brexit-Related Uncertainty Index (BRUI). In creating this index, we apply Text Mining, Context Window, Natural Language Processing (NLP), and Large Language Models (LLMs) from Deep Learning techniques to analyse the monthly country reports of the Economist Intelligence Unit from May 2012 to January 2025. Additionally, we employ a standard vector autoregression (VAR) analysis to examine the model-implied responses of various macroeconomic variables to BRUI shocks. While developing the BRUI, we also create a complementary COVID-19 Related Uncertainty Index (CRUI) to distinguish the uncertainties stemming from these distinct events. Empirical findings and comparisons of BRUI with other earlier-developed uncertainty indexes demonstrate the robustness of the new index. This new index can assist British policymakers in measuring and understanding the impacts of Brexit-related uncertainties, enabling more effective policy formulation.

## 1. Introduction

One of European history's most significant political and economic ruptures is the UK's withdrawal from the European Union (EU), commonly known as Brexit. This withdrawal has had considerable effects not only on the UK and the EU but also on other trade partners globally (Brakman et al., 2018; Anderson & Wittwer, 2018; Kren & Lawless, 2024; Buigut & Kapar, 2025). Since the referendum in 2016, the uncertainties brought by Brexit have affected many areas, such as financial markets, investment decisions, export/imports, exchange rates, and supply chains (Panitz & Glückler, 2022; Hassan et al., 2024; Du et al., 2025; Ongan et al., 2025a; Ongan et al., 2025b). Renegotiation of trade agreements and restrictions on labour mobility have further complicated the economic effects of Brexit (Forslid & Nyberg, 2021; Sargent, 2023; Cusimano et al., 2024; Du et al., 2025).

Even though Brexit is now formally complete, its economic consequences continue to materialize, making tools that precisely measure its associated uncertainties valuable for both practical policymaking and academic understanding. The UK is still adapting to new trade relationships with the EU (Michail, 2021), new regulatory frameworks, and labour market changes and is continuing to negotiate potential new trading arrangements with a range of non-EU countries/regions.

While various approaches to measuring Brexit-related uncertainties exist in the literature, as will be discussed in subsequent sections, they suffer from critical limitations: most have not been updated to reflect the evolution of uncertainty during the Brexit process, and importantly, they fail to disentangle Brexit-related uncertainties from those stemming from the COVID-19 pandemic, creating a significant gap in the availability of a comprehensive and precisely targeted metric that directly captures Brexit-specific economic uncertainties throughout the withdrawal process and the immediate aftermath. This study addresses these gaps by developing and introducing the Brexit-Related Uncertainty Index (BRUI), which offers an up-to-date measurement of Brexit-specific uncertainties. Crucially, since COVID-19 emerged as a significant concurrent phenomenon during our sample period, we have also developed a complementary COVID-19 Related Uncertainty Index (CRUI). This additional index serves as both a methodological counterpart and a statistical control, enabling us to effectively disentangle Brexit-specific uncertainties from pandemic-induced economic disruptions—a distinction absent in existing measures. This methodological innovation is particularly important because the pandemic's widespread economic effects could otherwise obscure or conflate the distinct impacts of Brexit on the UK economy.

The contributions of the study and the advantage of using this index to the literature can be listed as follows:

- BRUI is a new composite index created to measure uncertainties related to Brexit. It allows for a more accurate analysis of the uncertainties arising from this phenomenon.

- By quantifying Brexit-related uncertainty objectively and systematically, BRUI enables a more rigorous assessment of Brexit's economic impacts, distinguishing them from other contemporaneous economic shocks and enabling better-informed policymaking.

- The index can be used as either an independent or dependent variable in time in macroeconomic time series models, enhancing analytical capabilities in econometric research.

- Unlike studies that represent Brexit as a static dummy variable, this index captures its dynamic nature, tracking how Brexit-related uncertainty evolved over time in intensity and character—from anticipatory uncertainty through negotiation phases and on to implementation uncertainties.

An increase in BRUI indicates that Brexit-related uncertainty is rising, potentially leading to heightened market volatility, reduced business confidence, delayed or diverted investment decisions, and disruptions to trade and other economic activities.

The remainder of the study is organized as follows: Sections 2 and 3 present a theoretical background and literature review. Section 4 outlines the empirical methodology and models, and section 5 presents the empirical findings. Finally, Section 6 discusses the conclusions with policy implications, and Section 7 presents the study limitations with future research recommendations.

## 2. Theoretical background

This study's theoretical background is based on the following theories and frameworks relating the Brexit process to uncertainty.

The *New Institutional Economics Theory*, as outlined by North (1990), examines the role of institutions in shaping economic activity. This framework integrates economics with other social sciences, such as political science and law, to provide a comprehensive analytical perspective. Brexit has triggered a significant institutional transformation through new regulations, trade agreements, and legal frameworks. Thus, this theory is highly relevant for analyzing Brexit, as it underscores the profound effects of institutional changes—particularly regarding property rights, transaction costs, and the disruption of established economic relationships within the European Union. In this context, BRUI can serve as a valuable quantitative indicator (metric) for assessing the impact of Brexit-induced institutional uncertainty on economic growth, trade, and investment decisions.

The *Trade Policy Uncertainty (TPU)* framework, examined by Handley & Limão (2015), explains how uncertainty about future trade policies—such as tariff changes or regulatory shifts—affects firms' investment and export decisions. This framework is particularly relevant to the Brexit process, as the UK's trade relations with the European Union (EU) have been marked by significant uncertainty. Therefore, BRUI will serve as a crucial tool for measuring the extent to which the UK economy is impacted by trade policy uncertainty. Baker et al. (2016) transformed trade policy uncertainty into an index for the United States. If this index were developed for the UK, it would provide both a theoretical and practical foundation for our effort to quantify BRUI in this study. Thus, this index would allow us to systematically compare the BRUI and TPU at the index level.

## 3. Literature review

The literature features various studies investigating the effects of Brexit-related uncertainties on different aspects of the economy. For instance, certain studies have concentrated on the impacts of these uncertainties on international financial markets (Smales, 2017; Belke et al., 2018; Kellard et al., 2022; Rodella et al., 2023; Koch et al., 2024). Others have examined how Brexit uncertainties affect international trade (Graziano et al., 2018; Matzner et al., 2023; Du et al., 2025). Additionally, some research has delved into the repercussions of Brexit-related uncertainties on firms (Vasilescu & Weir, 2023; Hassan et al., 2024). Furthermore, studies have explored the effects of these uncertainties on economic activities (Biljanovska et al., 2017) and investments (Meinen and Röhe, 2017).

Several methodological approaches have been developed in the literature to measure Brexit uncertainty. Hassan et al. (2020) proposed a comprehensive text-based approach utilizing Natural Language Processing (NLP) to assess the effects of uncertainties associated with Brexit. Alternative measurement strategies include Graziano et al. (2018), who analysed monthly export variations, and Handley et al. (2020), who employed prediction market-based variables. Steinberg (2019) used a dynamic model to explore the effects of increased trade costs resulting from Brexit. Event studies were the preferred methodology for Oehler et al. (2017) and Ramiah et al. (2017), while Belke et al. (2018) focused on policy uncertainty, specifically utilizing the Economic Policy Uncertainty (EPU) Index. Schiereck et al. (2016) investigated the reactions of equity and CDS investors to the referendum announcement. More recently, Makrychoriti & Spyrou (2023) analysed the international economic effects of Brexit uncertainty using existing measures such as the EPU Index and Brexit dummy variables, but without developing new metrics that distinguish Brexit effects from other concurrent economic shocks.

Several studies have developed specialized indices to measure Brexit-related uncertainty. Baker et al. (2016) developed the Brexit-related uncertainty index based on their economic policy uncertainty (EPU) methodology. Their index, abbreviated in this study as BRUI_Baker, was constructed by rescaling the general EPU to isolate Brexit-related uncertainty, analysing the frequency of specific keywords (economy, policy, uncertainty, tax, regulation, Brexit, EU) in major British newspapers, including The Financial Times and The Times of London. However, their index covers only 2000-2016, so it is suited to capturing the impacts of uncertainty about the referendum result and avoids any concerns about separating Brexit and COVID-19-related uncertainties, but it excludes the critical post-referendum negotiations and eventual implementation period. Another disadvantage is that newspaper-based analysis introduces potential editorial biases in uncertainty measurement.

Bloom et al. (2019) created another Brexit-related uncertainty index, abbreviated in this study as BRUI_B. Developed with Bank of England support, this index employed the Bank's Decision-Making Panel (DMP) survey to capture firms' perceptions of Brexit-related uncertainty between January 2015 and April 2024. The index was specifically designed to understand how businesses formed expectations and made investment decisions during the Brexit process. The index spans the period from January 2015 to April 2024. Although this method helps provide a real-time measure of uncertainty based on firms' direct perceptions and responses, its reliance on subjective assessments may cause perceptions to deviate from economic realities. Additionally, not including the uncertainty perceptions of other economic actors (consumers, financial markets, etc.) may narrow the relevance of the index. One notable limitation is that there is no accompanying measure of uncertainties related to COVID-19, so separate identification of impacts of Brexit- and COVID-19-related uncertainties is not feasible.

The final alternative is the Brexit Uncertainty Index (BRUI_C) developed by Chung et al. (2022). This index employed advanced Natural Language Processing techniques, specifically Continuous Bag-of-Words (CBOW) architecture, to analyse news articles from leading UK publications. Covering 2013-2022, it measured aggregate Brexit and disaggregated uncertainty across specific policy domains, including trade, immigration, Northern Ireland, supply chains, energy, and employment. Although BRUI_C introduced a method to separate COVID-19 and Brexit-related uncertainty, it assumes that uncertainty was driven solely by COVID-19, excluding other potential factors. Furthermore, BRUI_C has not been updated beyond 2022.

In contrast, our study introduces the Brexit-Related Uncertainty Index (BRUI), which offers a more sophisticated methodology by complementing and enhancing the studies above. The BRUI utilizes text mining, Natural Language Processing (NLP), and large language models (LLMs) from Deep Learning, which can also distinguish COVID-19-related uncertainties. BRUI allows the level of Brexit-related uncertainty to change dynamically each month, covers the period in which Brexit and COVID-19-related uncertainties were prevalent, and has taken steps to disentangle these influences on uncertainty.

    i)    The newly developed BRUI offers significant methodological improvements and practical benefits that address the limitations of previous indices. Notably, it utilizes standardized source material by employing the Economist Intelligence Unit's (EIU) monthly country reports. This ensures consistent comparability of values over time due to the uniform reporting formats used in these reports.

ii) Comprehensive Temporal Coverage: This extends beyond the EU referendum through the negotiation and implementation of Brexit and into the post-Brexit period. It can be updated monthly, and researchers will have free access.

iii) Methodological Separation of Concurrent Factors: Effectively decomposes uncertainties arising from Brexit and COVID-19, avoiding the conflation of impacts of these distinct economic shocks.

iv) Dynamic Weighting System: Implements a dynamic weighting approach rather than applying static ratios for uncertainties simultaneously affected by COVID-19 and Brexit.

v) Contextual Analysis: Goes beyond simple keyword identification by requiring uncertainty terms to appear together with Brexit references within the same context window, enhancing measurement precision.

vi) Avoidance of ambiguity around Referendums: Carefully distinguishes between different referendum discussions by excluding context windows containing "referendum" alongside "Scotland" or "Scottish," preventing conflation with Scottish independence discussions.

vii) Uses advanced analytical techniques: Combining context window (CW) techniques, text mining with Natural Language Processing (NLP), and Large Language Models (LLMs) to create a more nuanced differentiation between Brexit and COVID-19 generated uncertainties.

As a complementary contribution, the study also develops a COVID-19 Related Uncertainty Index (CRUI) for the UK, which will be made freely available to researchers, enabling more precise analysis of pandemic-specific economic impacts.

## 4. Empirical methodology and models

The following steps were followed in developing the new Brexit-Related Uncertainty Index (BRUI) in this study:

*Step 1:* This study's sample period is between May 2012 and January 2025 since the term "Brexit" was first used in the press in May 2012 (Chung et al., 2023).

*Step 2:* The keywords in Table 1 related to uncertainty, Brexit, and COVID-19 were identified by following the methodologies of Baker et al. (2016a, 2016b), Ferreira et al. (2019), Ahir et al. (2022),

Chung et al. (2023), and Dang et al. (2023) and the Web of Science (WOS) data set. These keywords were then scanned in Economic Intelligence Unit (EIU) reports.

**Table 1. Uncertainty, Brexit, and COVID-19 related keywords**

| Category | Keywords |
|---|---|
| Uncertainty Related | fear, indecision, instability, jittery, nervousness, precarious, tense, tension, uncertain, uncertainly, uncertainty, unclear, unknown, unpredictable, unsettled, unstable, volatile, volatility, worry |
| Brexit Related | article 50, Brexit, Brexit-related, customs union, EU exit, EU membership, EU withdrawal, exit deal, exit from the EU, exit the EU, exit time, exiting, exiting the EU, exiting the European union, free movement, internal market bill, leave the EU, northern Ireland protocol, post-Brexit, pre-Brexit, referendum, regulatory alignment, regulatory framework, single market, trade arrangement, trade negotiations, transition period, UK exits, UK-EU relations, Uk-EU trade deal, UK's withdrawal, withdrawal agreement, withdrawal from the EU |
| COVID-19 Related | coronavirus, covid, covid-19, lockdown, outbreak, pandemic, quarantine, vaccination, vaccine |

Note 1: Exact matches to the specified keywords/phrases were identified in the text using NLP's "n-gram" method. Note 2: In EIU reports on Brexit, the term "referendum" appears with various prepositions and words. To avoid confusion with the 2014 Scottish independence referendum, any context containing "referendum" alongside "Scotland" or "Scottish" was excluded from the analysis and omitted from calculations. Note 3: The term "COVID" appeared in the reports also without "19," so "COVID" was added to the keyword list to ensure all instances were captured. Note 4: In text mining, NLP, and LLM analyses, all words were converted to lowercase before processing. Note 5: In the initial phase, a larger set of keywords was used. Then, a Python script identified the frequency of each keyword, and those absent in EIU reports were removed from the list.

*Step 3:* The Fitz module from Python's PyMuPDF library made report texts searchable in PDF format. Manual verification confirmed its successful performance, with all text converted to lowercase.

*Step 4:* The text was tokenized into words using the Natural Language Toolkit (NLTK) module in NLP. The NLTK offers a comprehensive set of tools for NLP, covering tasks such as tokenization, sentence parsing, and stemming (Upreti, 2023).

*Step 5:* The 'stopwords' module of NLTK removed stop words (such as 'a,' 'is,' and 'the').

*Step 6:* "*n-grams"* used to analyse word sequences, examining bi-grams (e.g., "*customs union*") and three-grams (e.g., "*exit the EU*"). An n-gram represents a sequence of *n* consecutive words and can include more than three words depending on *its* value.

*Step 7:* The Named Entity Recognition (NER) tool considered other words related to the given keywords. Thus, we counted total word numbers and total uncertainty word numbers on monthly reports.

*Step 8:* This study employs SpaCy's "*en_core_web_lg*" module of Large Language Models (LLMs) to elucidate the contextual occurrence of "*uncertainty words (U)*" ascertaining their association with Brexit or COVID-19. SpaCy, an LLM-based NLP library, uses Named Entity Recognition (NER) to identify entities and analyse grammatical relationships, aiding tasks like sentiment analysis and spam detection (Domino, 2024). It effectively enables complex NLP processes and extracts key insights from text data.

*Step 9:* To identify the associations between "uncertainty-related words" and "Brexit and COVID-19 related words", we used the following Context Windows *(CW)* technique, which includes 10 words before and after each uncertainty keyword (U). Regarding the rationale for selecting a 10-word range, we experimented with different window sizes and manually examined reports to assess how far keywords appear before or after U, ultimately determining that a 10-word range was the most appropriate based on these observations:

$$CW = \{x_{-10}, \ldots, x_{-2}, x_{-1}, U, x_{+1}, x_{+2}, \ldots, x_{+10}\} \qquad (1)$$

where U represents an uncertainty-related keyword, and $x_{-i}$ and $x_{+i}$ denote any keywords appearing before and after U. If a context window (CW) contains a Brexit-related keyword (BRK) and does not include any COVID-19-related keyword (CRK), the uncertainty is classified as *Brexit-related uncertainty*, and the count of 'Brexit-Related Uncertainty Keyword Number (BRUKN)' is increased by one, as shown in the following form:

$$BRUKN = \begin{cases} BRUKN + 1, & \text{if a BRK exists in CW and not exists any CRK in CW} \\ BRUKN, & \text{otherwise} \end{cases} \qquad (2)$$

*Step 10:* If a context window includes the Brexit-Related Keyword (BRK) and COVID-19-Related Keyword (CRK) together, they were classified as "Brexit & COVID-19-Related Uncertainty", then using proportion (weight) of "Brexit-Related Uncertainty Word Number" among "Brexit-Related

Uncertainty Keyword Number" + "COVID-19-Related Uncertainty Keyword Number", the Total Brexit-Related Uncertainty Word numbers (TBRUKN) were obtained. We apply an iterative mechanism using a sliding window approach.

To consider the varying page lengths and word counts in monthly reports, we implement a standardization process based on the method described by Ahir et al. (2022). This involves calculating the ratio of the number of identified Total Brexit Related Uncertainty Keyword Numbers ($TBRUKN$) in each report to the total word count of that report in the following form:

$$BRUI_t = \frac{TBRUKN_t}{(Total\ Number\ of\ Words\ Per\ Report)_t} \qquad (3)$$

where $BRUI$ represents the standardized measure of Brexit-Related Uncertainty, $TBRUKN$ denotes the count of Brexit-Related Uncertainty Keywords, and the $Total\ Number\ of\ Words\ Per\ Report$ accounts for variations in report lengths. The variable *t* represents the month of the EIU report.

*Step 11:* Then, the index was normalized to ensure its maximum value of *BRUI* is 100, following the methods described by Dang et al. (2023) and Chung et al. (2023). This normalization facilitates comparison across reports and time periods. A higher value of the BRUI reflects greater Brexit-related uncertainty, while a lower value indicates less uncertainty. Figure 1 illustrates the above context window (*CW*) methodology in a simplified form:

**Figure 1. Simplified presentation of the methodological process.**

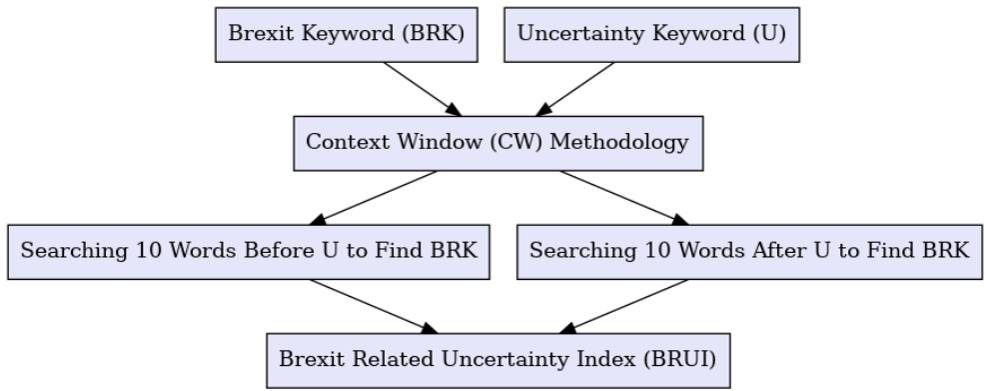

Source: Created using Python (Graphviz library).

To explain the methodology presented in Figure 1 with a concrete example, we first catch the use of an uncertainty-related keyword (*U*) like "uncertainty," then check for the presence of any Brexit-related keywords, like "Exit the EU," within 10 words to the right and left of *U*. If we met a Brexit related keyword, this word classified as Brexit Related Uncertainty Keyword (BRUK). To obtain the Brexit Related Uncertainty Index (BRUI), the Total BRUK Number (TBRUKN) is divided by Total Word Numbers in the Related Report of EIU for standardization. Finally, we normalized BRUI as a max value of 100. A higher value in the normalized index indicates greater uncertainty and vice versa. We also created a COVID-19 Related Uncertainty Index (CRUI) following this system.

## 5. Empirical findings

In this section, Figure 2 first presents the three-period BRUI and major global events throughout the study period.

**Figure 2. The BRUI and major global events.**

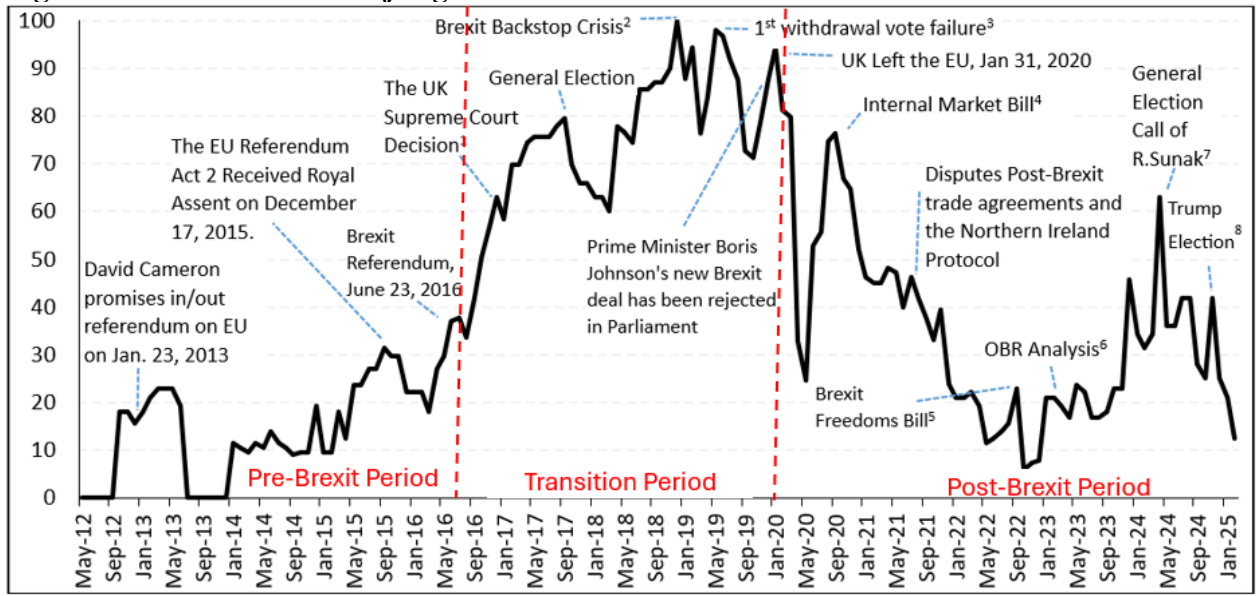

Source: Prepared by the authors.

As shown in Figure 2, uncertainties related to Brexit started to rise when discussions began in January 2013. This trend intensified following the 'yes' outcome of the June 2016 referendum. Market uncertainties continued to increase and fluctuate, a pattern attributed to the mismanagement of the Brexit process. As noted by Ward (2021), this situation was further compounded by David Cameron's resignation as Prime Minister on the morning after the referendum.

Focusing on crucial events in Figure 2 that impacted Brexit-related uncertainty: (1) In November 2016, the UK Supreme Court ruled that parliamentary approval was required to trigger Article 50, heightening concerns over a prolonged Brexit process. (2) The 2018-2019 "Irish backstop" issue created deep political divisions and uncertainty as the UK struggled to pass an agreement on the Northern Ireland border. (3) Theresa May's Withdrawal Agreement was defeated in Parliament on January 15, 2019, marking a historic government loss and escalating Brexit-related uncertainties. (4) In September 2020, the UK introduced the Internal Market Bill, suggesting possible protocol violations, particularly regarding Northern Ireland, further straining EU-UK relations. (5) On September 22, 2022, the "Brexit Freedoms Bill" aimed to eliminate EU laws by 2023, creating additional uncertainty over regulatory changes, while economic policies by Liz Truss's government sparked market instability. (6) The Office for Budget Responsibility projected a 4% long-term drop in UK productivity due to post-Brexit trade frictions, with an expected 15% decrease in trade. (7) Lastly, Trump's re-election as U.S. President for a second term in November 2024 has led to a renewed increase in Brexit-related uncertainties in the UK.

It can be thought that the key policy issues related to Brexit have been resolved and that uncertainty is no longer a determining factor for business and policy decisions. However, uncertainties may still exist in the post-Brexit period. In particular, trade agreements with non-EU countries, financial regulations, and immigration policies may continue to create uncertainty for businesses. Moreover, although the effects of Brexit are considered mainly retrospective, this uncertainty may resurface during election periods and economic fluctuations.

Moreover, the possibility of the UK rejoining the EU (which we may define as Brin) should not be overlooked, as the process triggered by the Russia-Ukraine war, the U.S.'s 2025 high tariff policies, Europe's deteriorating economic conditions, migration crises, and shifts in energy policies may increase uncertainties, making reunifications and new restructurings inevitable.

To test the robustness of the BRUI, we compare the other indices related to Brexit uncertainties in Figures 3, 4, and 5. First, Figure 3 compares the BRUI with Bloom et al. (2019)' Brexit-related uncertainty index (BRUI_B).

**Figure 3. Comparison of BRUI with BRUI_B.**

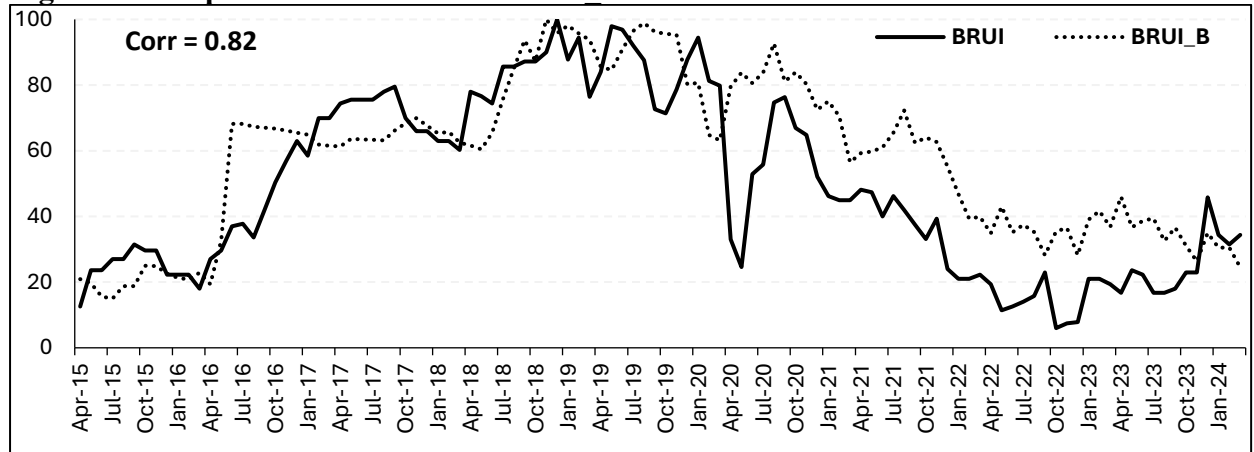

Source: BRUI was prepared by the authors. BRUI_B index series were obtained from the Bank of England (2025).

Figure 3 shows a high correlation of 0.82 (denotes the robustness of the BRUI) between BRUI and BRUI_B, providing strong evidence that BRUI effectively identifies Brexit-related uncertainties in the UK. The discrepancies between BRUI and BRUI_B are attributable to differences in data sources. Specifically, while the Bank of England (2025) relied on data that reflects responses to their Decision Maker Panel survey, our study utilizes reports from the Economist Intelligence Unit (EIU). Figure 4 compares the BRUI with Chung et al. (2023)'s Brexit Uncertainty index (BRUI_C).

**Figure 4. Comparison of BRUI with BRUI_C.**

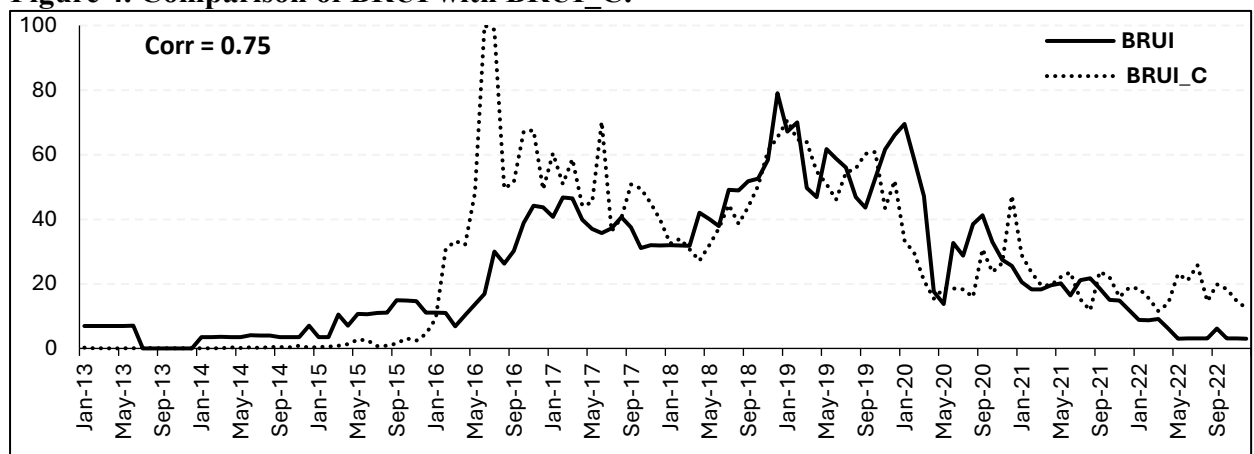

Source: BRUI was prepared by the authors and BRUI_C was Chung et al. (2023).

Despite differences in data sources, the high correlation of 0.75 (denotes the robustness of BRUI) between BRUI and the BRUI_C developed by Chung et al. (2023) highlights the effectiveness of BRUI in detecting Brexit-related uncertainty in the UK. These variations between BRUI and BRUI_C are attributed to differences in keywords and data sources. Specifically, Chung et al. (2023) relied on newspaper data, whereas this study utilized reports from the EIU.

The smaller differences observed in 2021 and 2022 compared to 2016 may be due to the evolving nature of Brexit uncertainty over time. While the 2016 referendum process led to sudden and significant fluctuations, Brexit uncertainty during the COVID period may have been more intertwined with broader economic factors, potentially reducing its relative impact on the index. However, the high correlation (0.75) between BRUI and BRUI_C confirms that Brexit-related uncertainty is reliably and effectively captured. Therefore, the change in the magnitude of observed differences could be interpreted as a natural outcome of the evolution of uncertainty over time rather than a methodological shortcoming.

Lastly, Figure 5 compares the BRUI with Baker et al. (2016)' Brexit-related uncertainty (BRUI_Baker).

**Figure 5. Comparison of BRUI with BRUI_Baker.**

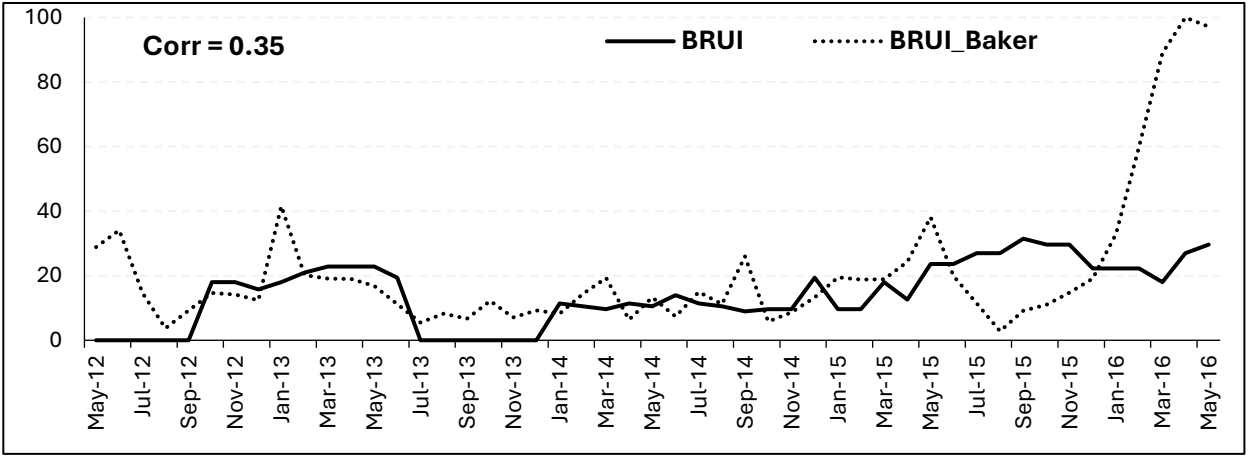

Source: BRUI was prepared by the authors, and BRUI_Baker was Baker et al. (2016).

The BRUI_Baker index ended at the referendum on EU membership in June 2016, while BRUI continues to provide a dynamic measurement covering the post-referendum period. Comparison between the indices is limited to data up to June 2016. As explained previously, the BRUI_Baker index is derived from analysis of newspaper articles, while the BRUI index uses EIU reports; there are also some differences in the methodologies used. Despite these differences, both indices moved in parallel until 2016, indicating that pre-Brexit uncertainty was similarly perceived across different sources. The similar trends in both indices suggest BRUI's partial robustness.

The difference in uncertainty levels around the EU referendum in June 2016 may be attributed to the nature of the data sources used. While BRUI_C and BRUI_Baker rely on daily newspaper reports, BRUI is constructed using monthly EIU reports. Daily newspaper data captures short-term fluctuations and immediate reactions to political events, potentially amplifying uncertainty spikes. In contrast, monthly reports provide a more structured and aggregated assessment, smoothing transient volatility and focusing on sustained uncertainty trends. This methodological distinction explains why BRUI may exhibit a relatively lower peak during the referendum period while still effectively capturing Brexit-related uncertainty over time.

Next, we will conduct standard VAR (vector autoregression) analysis using US data for 2012M5-2025M1 to examine the model-implied responses of some macroeconomic variables to shocks and Brexit-related uncertainty. These macroeconomic variables we consider are Gross Domestic Product (GDP), Consumer Price Index (CPI), Producer Price Index (PPI), Exports (X), Imports (M), the British Pound to Euro exchange rate (GBP_EUR), the British Pound to USD exchange rate (GBP_USD), Employment (EMP), and the Unemployment Rate (UEMP). Table 2 summarizes the variables and their codes, definitions, and sources. The impulse-response functions are shown in Figure 6.

**Table 2. Selected UK monthly macroeconomic variables.**

| Variable Codes | Variable Definitions | Sources |
|---|---|---|
| **Uncertainty-related Variable** | | |
| BRUI | Brexit Related Uncertainty Index for the UK | Created by the authors |
| **Macroeconomic Variables** | | |
| GDP | Monthly GDP Index (2022=100) | ONS (2025a) |
| CPI | Consumer Price Index (2015 = 100) | ONS (2025b) |
| PPI | Producer Price Index (2015 = 100) | ONS (2025c) |
| X | Trade in Goods (Billion Pounds, £) | ONS (2025d) |
| M | Trade in Goods (Billion Pounds, £) | ONS (2025e) |
| GBP_EUR | GBP/EUR - British Pound Sterling Euro exchange rate | ONS (2025f) |
| GBP_USD | GBP/USD - British Pound Sterling US dollar exchange rate | ONS (2025g) |
| EMP | Number of People in Employment (aged 16 and over, million) | ONS (2025g) |
| UEMP | Unemployment rate (aged 16 and over, %) | ONS (2025h) |

**Note:** All variables (except GBP_EUR, GBP_USD, and UEMP) are expressed in logs. Because GBP_EUR, GBP_USD, and UEMP have small values, and since all variables are non-stationary at the level, we used the first differences of all series in the VAR analyses.

In the VAR analyses, impulse-response functions were used to examine the responses of macroeconomic variables to a one-standard-deviation positive shock in BRUI. The variance decomposition analysis revealed the impact of changes in BRUI on macroeconomic variablesFigure 6 presents the impulse-response functions based on the VAR analysis. The impulse-response functions were estimated using standard VAR analysis, with 90% confidence intervals and standard percentile bootstrap, employing 999 bootstrap repetitions.

The impulse-response charts presented below indicate that a one standard deviation positive shock to Brexit-related uncertainty results in a decline in GDP, PPI, X, M, GBP_EUR, and EMP, while CPI, GBP_USD, and UEMP exhibit an increase. The contraction in imports is more pronounced than that in exports. CPI responds immediately to BRUI, whereas PPI's upward response becomes more significant over the long term. Employment has declined as many EU citizens who worked in the UK subsequently left, citing Brexit as the reason (Aerssen and Spital, 2023).

The impulse-response functions indicate that, following a one-standard-deviation positive shock to Brexit-related uncertainty, the British Pound initially depreciated against both the Euro and the USD despite showing a short-term appreciation against the USD. Over time, the Pound exhibited a partial recovery against the USD, yet it did not fully return to its pre-shock levels. These responses suggest that Brexit-related uncertainty had a significant but not entirely persistent impact on exchange rates.

Considering the impact on PPI, the sharp decline in GDP, the decrease in EMP, and the increase in UEMP, BRUI has significantly negatively affected production in the UK.

**Figure 6. Responses of macroeconomic variables to a shock in the BRUI.**

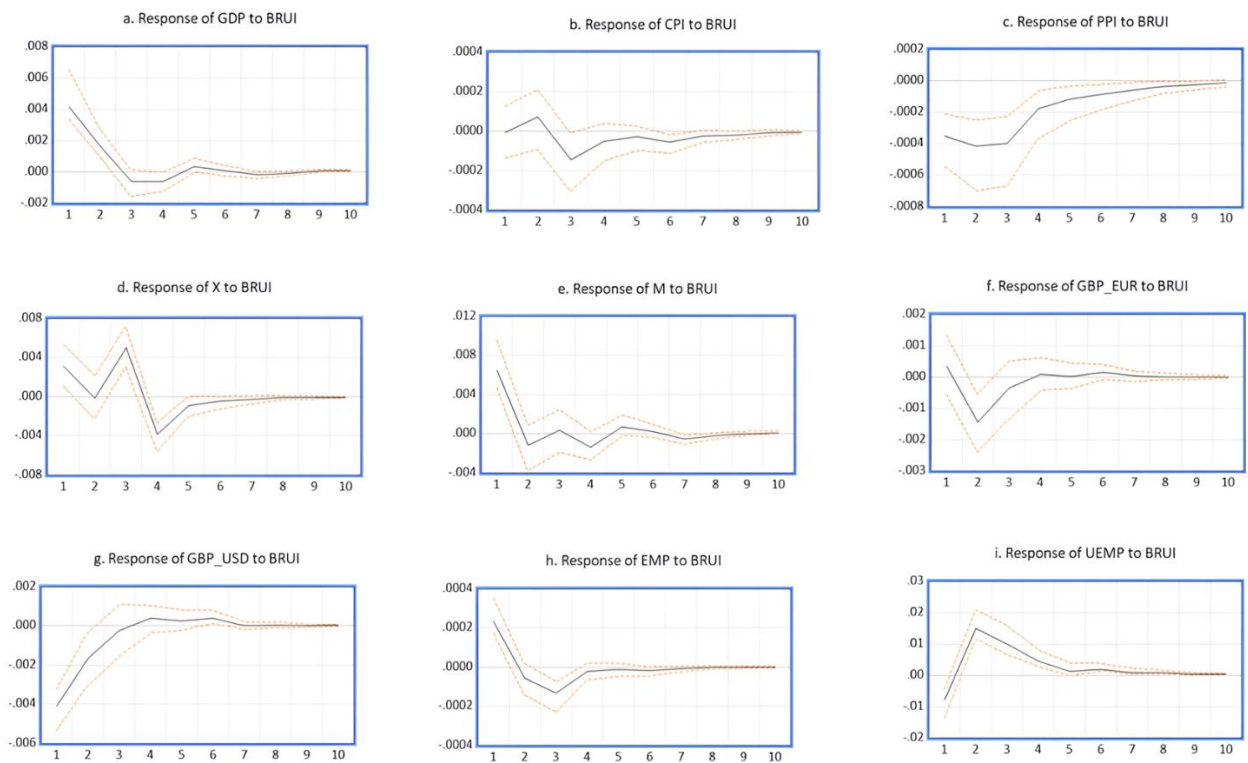

The Cholesky forecast-error variance decomposition (FEVD) was also conducted to determine how much a variable's forecast error variance is attributable to shocks from other variables. The FEVD method is commonly used in time series analysis, particularly within the context of VAR models (Ellington, 2018; Albert & Agnes, 2024). Table 3 presents the explanatory power of a one-standard-deviation change in BRUI on macroeconomic variables.

**Table 3. Cholesky forecast-error variance decomposition results.**

| Period | BRUI | GDP | CPI | PPI | X | M | GBP_EUR | GBP_USD | EMP | UEMP |
|---|---|---|---|---|---|---|---|---|---|---|
| 1 | 100 | 0 | 0 | 0 | 0 | 0 | 0 | 0 | 0 | 0 |
| 2 | 90.72 | 2.15 | 0.00 | 0.85 | 0.04 | 1.66 | 0.52 | 4.02 | 0.03 | 0.00 |
| 3 | 85.91 | 3.02 | 0.20 | 0.81 | 0.06 | 2.44 | 2.44 | 4.44 | 0.05 | 0.64 |
| 4 | 85.24 | 3.33 | 0.25 | 0.85 | 0.15 | 2.47 | 2.42 | 4.42 | 0.25 | 0.63 |
| 5 | 84.82 | 3.33 | 0.25 | 0.96 | 0.29 | 2.46 | 2.43 | 4.43 | 0.29 | 0.73 |
| 6 | 84.69 | 3.39 | 0.29 | 0.98 | 0.31 | 2.46 | 2.43 | 4.42 | 0.30 | 0.73 |
| 7 | 84.63 | 3.39 | 0.29 | 0.99 | 0.32 | 2.48 | 2.43 | 4.42 | 0.33 | 0.74 |
| 8 | 84.60 | 3.40 | 0.29 | 0.99 | 0.32 | 2.49 | 2.43 | 4.41 | 0.33 | 0.74 |
| 9 | 84.59 | 3.40 | 0.29 | 0.99 | 0.32 | 2.49 | 2.43 | 4.41 | 0.33 | 0.74 |
| 10 | 84.58 | 3.40 | 0.29 | 0.99 | 0.33 | 2.49 | 2.43 | 4.41 | 0.34 | 0.74 |

The Cholesky forecast-error variance decompositions (FEVDs) in Table 3 show that since changes in BRUI stabilize after the 5th period, variance decomposition for this period has been examined. The results indicate that fluctuations in BRUI have had the most significant impacts on GDP, the value of the British Pound against the USD (GBP_USD), imports (M), and the value of the British Pound against the Euro (GBP_EUR). These results reveal that Brexit-related uncertainty has directly affected UK economic growth and exchange rates.

Additionally, this analysis confirms that BRUI's impact on PPI is greater than that of CPI. A possible explanation for this result is that companies' production costs may be more sensitive to Brexit-related uncertainties than consumer prices. BRUI's effect on imports (M) is stronger than on exports (X). This may indicate that Brexit-related uncertainties make import processes more difficult, especially regarding foreign trade balance. Finally, BRUI has more pronounced effects on the unemployment rate (UEMP) than employment (EMP). Ultimately, the results indicate that Brexit-related uncertainty causes employment losses and a higher unemployment rate in the UK.

Moreover, in addition to the above evaluations, impulse-response functions and Cholesky forecast-error variance analyses were conducted in three separate phases—pre-Brexit, transition, and post-Brexit—and the results are presented in Figures 7, 8, and 9 and Tables 4, 5, and 6 in Appendixes 1, 2, and 3.

In Appendix 2, during the Brexit transition period (2016M07 – 2020M01), when BRUI experienced its sharpest increase, a one-standard-deviation shock to BRUI explains the changes in BRUI itself more significantly while explaining macroeconomic variables to a lesser extent. In other words, BRUI became a standalone issue during this period, influencing macroeconomic magnitudes rather than being affected by other macroeconomic variables. This effect can also be observed in the sharp fluctuations in the impulse-response graphs. The changes in BRUI have had a more aggressive impact on macroeconomic magnitudes.

On the other hand, in Appendix 3, during the post-Brexit period (2020M02 – 2025M01), a similar one-standard-deviation shock to BRUI explains the changes in BRUI itself more significantly while explaining macroeconomic variables to a lesser extent. In other words, during this period as well, BRUI continued to be a standalone issue. Rather than being influenced by other macroeconomic variables, BRUI itself influenced macroeconomic magnitudes. This effect can also be observed in the sharp fluctuations in the impulse-response graphs. The changes in BRUI have had a more aggressive impact on macroeconomic magnitudes.

## 6. Conclusions with policy implications

This study was motivated by the need to track the dynamic evolution of Brexit-related uncertainties and comprehensively analyse their effects on the UK macroeconomy. Utilizing The Economist Intelligence Unit (EIU) reports, Context Window analysis, Natural Language Processing (NLP), and Large Language Models (LLMs), our index, BRUI, specifically identifies the evolution of uncertainties attributable to Brexit.

Some studies reviewed above only account for Brexit using dummy variables, which cannot capture dynamic changes in Brexit-related uncertainties over time. As we discussed, among studies that capture changes over time, some rely on surveys reflecting a range of decision-makers perceptions, which may introduce subjective biases; others rely on newspaper reports and may lack the contextual analysis required to separate Brexit and COVID-19-related uncertainties.

Despite methodological differences, the high correlation coefficients from comparisons with other Brexit indices examined suggest that BRUI captures trends similar to those of other available measures.

Importantly, this new index (BRUI) covers both the pre-Brexit and post-Brexit periods starting from May 2012, when the Brexit concept was first used. The BRUI is available to researchers and policymakers upon request and will be updated as necessary.

BRUI's path over time clearly reveals the uncertainty dynamics of the Brexit process. There are three clear phases. The first phase began when the term Brexit was first used in March 2012 and covers the period in which Brexit uncertainty was largely over whether the UK would have a referendum on EU membership and whether the popular vote would be to leave the EU. The second phase began after the EU referendum result in June 2016, when uncertainty was mainly reflected in the lack of clarity and predictability around the Brexit process and future policy, including the immigration status of EU migrants in the UK and UK migrants living in EU countries, as well as the type of deal the UK would be able to negotiate with the EU, and any new trade deals with countries outside the EU. As the process continued and the 2020 deadline passed, a third phase was entered when policy issues associated with Brexit were resolved, the impacts of exit were realized, and a new normal came into being. However, the course of BRUI shows that references to Brexit-related uncertainties continue in reports, and arguably, the impact of Brexit is not entirely over.

Our findings allow us to conclude that Brexit constitutes a long-term source of structural uncertainty rather than a short-term shock effect. Ongoing fluctuations in BRUI continue to impact trade, inflation, and foreign exchange markets. Therefore, monitoring the BRUI provides a critical indicator for assessing the ongoing economic impacts of Brexit and developing new policy strategies for the UK.

In conclusion, by dynamically measuring Brexit-related uncertainties over time, BRUI can provide a valuable indicator (metric) for policymakers, investors, and businesses. Using BRUI, the impact of the UK's trade relations with the EU on supply chains can be examined, and proactive trade policies can be produced. BRUI can be used to analyse how UK-EU trade relations have changed post-Brexit. This new index can be used in academic research and econometric models to evaluate the structural transformations of the UK after leaving the EU. By following the BRUI, multinational companies and investors can foresee the risks associated with Brexit uncertainty and make more profitable investment decisions.

## 7. Limitations of this study and recommendations for future research

The BRUI developed in this study is limited to the keywords selected and the data source used. The index values may change with other keywords and data sources, so the number of keywords is kept as wide as possible, and Economic Intelligence Unit (EIU) reports are used instead of newspapers. Likewise, although the methodologies used in the study, context window, Natural Language Processing (NLP) techniques, and Large Language Models (LLMs), are powerful in detecting uncertainty and Brexit expressions, the possibility of using other methodologies can also be considered a limitation.

Therefore, future studies can use different data sources, create alternative keyword sets, and expand the scope of BRUI by applying different methodologies. Such studies can potentially increase the accuracy and reliability of the index, making a significant contribution to academic research and its use by policymakers. At the same time, integrating BRUI with measures of other sources of uncertainty will allow for a more in-depth analysis of economic impacts.

Additionally, future studies can conduct sectoral uncertainty analyses to examine the impact of BRUI on different sectors, such as finance, manufacturing, services, and agriculture, and compare how these sectors are affected by Brexit-related uncertainties. Likewise, they can examine the effects of BRUI on various macroeconomic variables, such as investment, international trade, GDP, exchange rates, and the labour force.

**Appendix 1. 2012M05 – 2016M06  Pre-Brexit Period.**

**Figure 7. Responses of macroeconomic variables to a shock in the BRUI.**

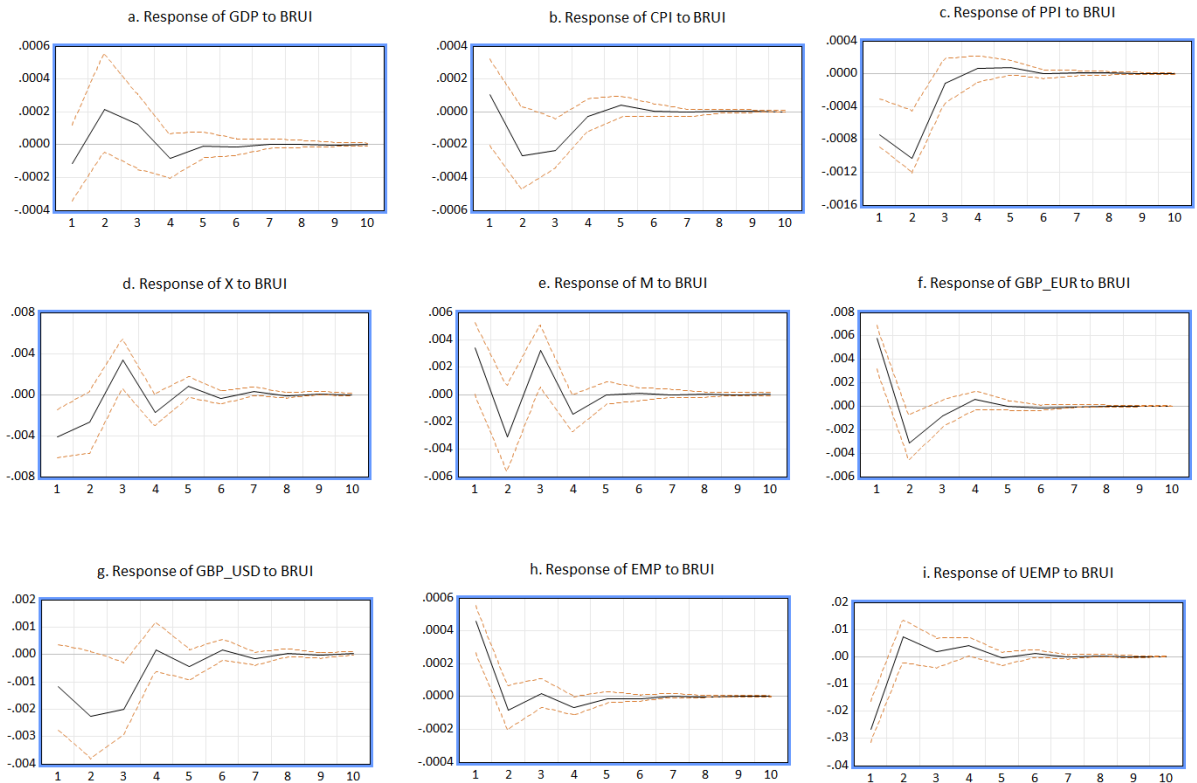

Table 4. Cholesky forecast-error variance decomposition results.

| Period | BRUI | GDP | CPI | PPI | X | M | GBP_EUR | GBP_USD | EMP | UEMP |
|---|---|---|---|---|---|---|---|---|---|---|
| 1 | 100 | 0 | 0 | 0 | 0 | 0 | 0 | 0 | 0 | 0 |
| 2 | 79.27 | 7.65 | 5.30 | 1.22 | 0.88 | 3.12 | 0.00 | 1.73 | 0.13 | 0.69 |
| 3 | 75.83 | 8.31 | 5.12 | 2.90 | 0.89 | 3.19 | 1.05 | 1.77 | 0.25 | 0.70 |
| 4 | 74.98 | 8.20 | 5.27 | 3.24 | 1.06 | 3.30 | 1.15 | 1.85 | 0.25 | 0.69 |
| 5 | 74.86 | 8.21 | 5.29 | 3.24 | 1.08 | 3.30 | 1.16 | 1.85 | 0.29 | 0.71 |
| 6 | 74.84 | 8.21 | 5.29 | 3.24 | 1.08 | 3.30 | 1.17 | 1.86 | 0.30 | 0.72 |
| 7 | 74.83 | 8.21 | 5.29 | 3.24 | 1.08 | 3.30 | 1.17 | 1.87 | 0.30 | 0.72 |
| 8 | 74.83 | 8.21 | 5.29 | 3.24 | 1.08 | 3.30 | 1.17 | 1.87 | 0.30 | 0.72 |
| 9 | 74.83 | 8.21 | 5.29 | 3.24 | 1.08 | 3.30 | 1.17 | 1.87 | 0.30 | 0.72 |
| 10 | 74.83 | 8.21 | 5.29 | 3.24 | 1.08 | 3.30 | 1.17 | 1.87 | 0.30 | 0.72 |

**Appendix 2. 2016M07 – 2020M01 Brexit Transition Period.**

**Figure 8. Responses of macroeconomic variables to a shock in the BRUI.**

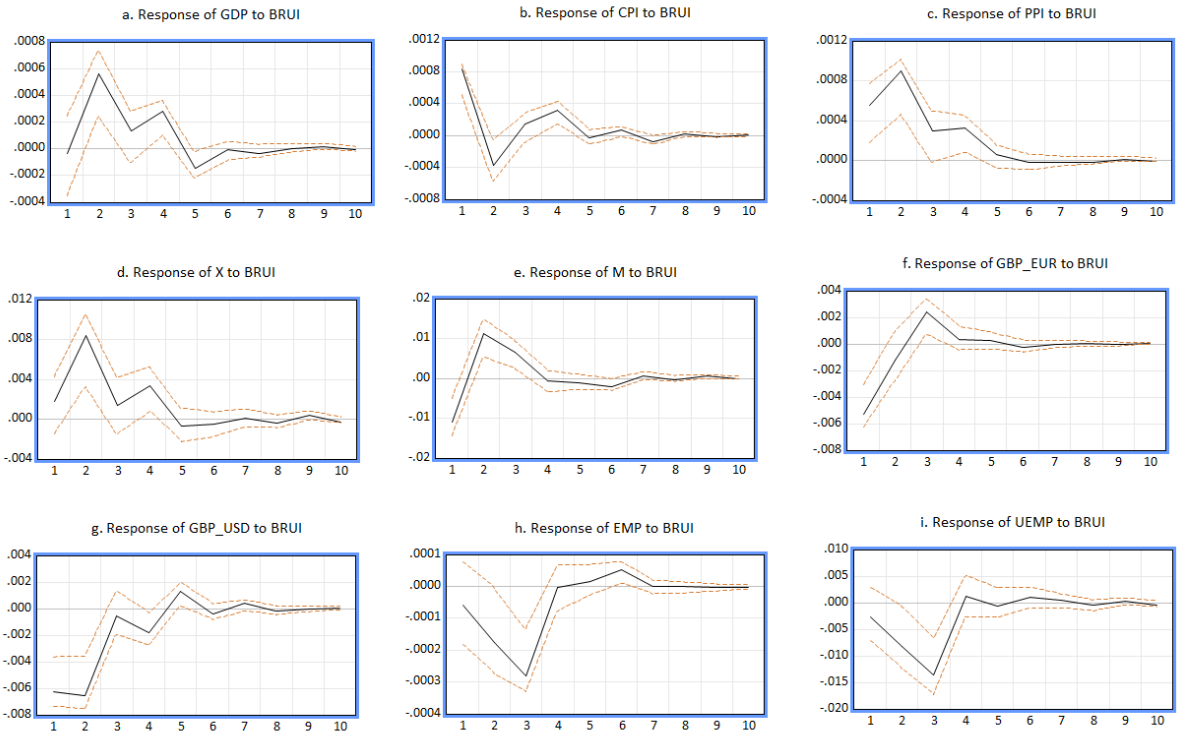

**Table 5. Cholesky forecast-error variance decomposition results.**

| Period | BRUI | GDP | CPI | PPI | X | M | GBP_EUR | GBP_USD | EMP | UEMP |
|---|---|---|---|---|---|---|---|---|---|---|
| 1 | 100 | 0 | 0 | 0 | 0 | 0 | 0 | 0 | 0 | 0 |
| 2 | 89.68 | 1.66 | 0.93 | 2.37 | 3.12 | 0.07 | 0.07 | 1.00 | 0.01 | 1.08 |
| 3 | 86.36 | 2.12 | 2.07 | 2.47 | 3.18 | 1.44 | 0.08 | 1.00 | 0.08 | 1.20 |
| 4 | 84.75 | 2.21 | 2.04 | 2.62 | 3.38 | 2.14 | 0.23 | 1.11 | 0.23 | 1.28 |
| 5 | 84.18 | 2.20 | 2.03 | 2.66 | 3.38 | 2.29 | 0.30 | 1.33 | 0.28 | 1.36 |
| 6 | 83.97 | 2.21 | 2.03 | 2.69 | 3.37 | 2.34 | 0.31 | 1.38 | 0.33 | 1.35 |
| 7 | 83.89 | 2.21 | 2.03 | 2.70 | 3.37 | 2.35 | 0.34 | 1.41 | 0.36 | 1.35 |
| 8 | 83.86 | 2.22 | 2.03 | 2.70 | 3.37 | 2.35 | 0.34 | 1.42 | 0.37 | 1.35 |
| 9 | 83.84 | 2.23 | 2.03 | 2.70 | 3.37 | 2.35 | 0.34 | 1.42 | 0.37 | 1.35 |
| 10 | 83.83 | 2.23 | 2.03 | 2.70 | 3.37 | 2.35 | 0.34 | 1.42 | 0.37 | 1.36 |

**Appendix 3. 2020M02 – 2025M01 Post-Brexit Period.**

**Figure 9. Responses of macroeconomic variables to a shock in the BRUI.**

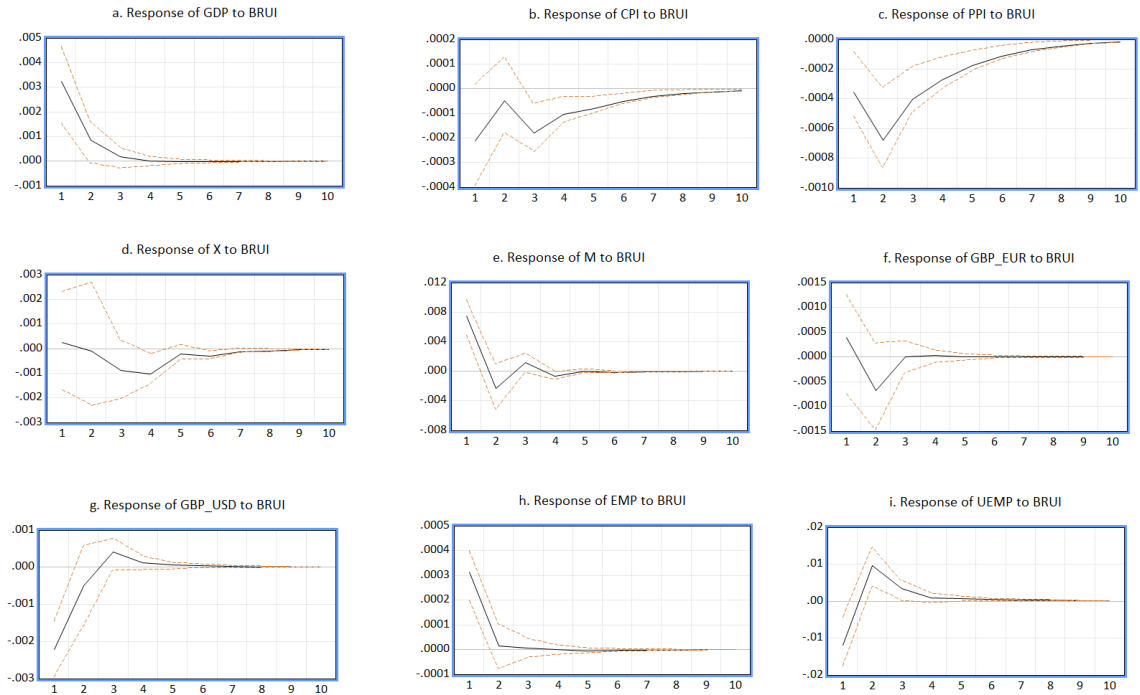

**Table 6. Cholesky forecast-error variance decomposition results.**

| Period | BRUI | GDP | CPI | PPI | X | M | GBP_EUR | GBP_USD | EMP | UEMP |
|---|---|---|---|---|---|---|---|---|---|---|
| 1 | 100 | 0 | 0 | 0 | 0 | 0 | 0 | 0 | 0 | 0 |
| 2 | 93.49 | 0.90 | 0.24 | 0.15 | 0.02 | 0.84 | 0.26 | 3.55 | 0.27 | 0.27 |
| 3 | 93.34 | 0.91 | 0.24 | 0.15 | 0.02 | 0.87 | 0.32 | 3.54 | 0.32 | 0.28 |
| 4 | 93.32 | 0.92 | 0.24 | 0.15 | 0.02 | 0.88 | 0.32 | 3.54 | 0.32 | 0.28 |
| 5 | 93.32 | 0.92 | 0.24 | 0.16 | 0.02 | 0.88 | 0.32 | 3.54 | 0.32 | 0.28 |
| 6 | 93.32 | 0.92 | 0.24 | 0.16 | 0.02 | 0.88 | 0.32 | 3.54 | 0.32 | 0.28 |
| 7 | 93.32 | 0.92 | 0.24 | 0.16 | 0.02 | 0.88 | 0.32 | 3.54 | 0.32 | 0.28 |
| 8 | 93.32 | 0.92 | 0.24 | 0.16 | 0.02 | 0.88 | 0.32 | 3.54 | 0.32 | 0.28 |
| 9 | 93.32 | 0.92 | 0.24 | 0.16 | 0.02 | 0.88 | 0.32 | 3.54 | 0.32 | 0.28 |
| 10 | 93.32 | 0.92 | 0.24 | 0.16 | 0.02 | 0.88 | 0.32 | 3.54 | 0.32 | 0.28 |